\documentclass{aastex702}
\usepackage{tikz}
\usepackage{hyperref}
\usepackage{orcidlink}
\usepackage{subfig}
\usepackage{newunicodechar}
\usepackage{graphicx}
\newunicodechar{₀}{$_0$}
\newunicodechar{−}{\textminus}

\begin{document}

\title{Likely Metal Scarring in the Magnetic White Dwarf LHS 2534}

\author[0009-0008-0863-1073]{Ryan J. Souchet}
\affiliation{Department of Astronomy, University of Florida, Bryant Space Science Center, Stadium Road, Gainesville, FL 32611, USA}
\email{r.souchet@ufl.edu}

\author[0000-0001-7143-0890]{Adam Moss}
\affiliation{Department of Astronomy, University of Florida, Bryant Space Science Center, Stadium Road, Gainesville, FL 32611, USA}
\email{drummeram@ufl.edu}

\author[0000-0002-4818-7885]{Jamie Tayar}
\affiliation{Department of Astronomy, University of Florida, Bryant Space Science Center, Stadium Road, Gainesville, FL 32611, USA}
\email{jtayar@ufl.edu}

\author[0000-0002-4462-2341]{Warren R. Brown}
\affiliation{Center for Astrophysics | Harvard \& Smithsonian,
60 Garden Street, Cambridge, MA 02138, USA}
\email{wbrown@cfa.harvard.edu}

\author[0000-0001-6098-2235]{Mukremin Kilic}
\affiliation{Homer L. Dodge Department of Physics and Astronomy, University of Oklahoma, 440 W. Brooks St., Norman, OK 73019, USA}
\email{kilic@ou.edu}

\author[0000-0001-5941-2286]{J.J. Hermes}
\affiliation{Department of Astronomy \& Institute for Astrophysical Research, Boston University, Boston, MA 02215, USA}
\email{jjhermes@bu.edu}

\begin{abstract}

There are only three white dwarfs known to possess metal-enriched polar caps due to magnetically guided accretion. These rare systems offer a unique opportunity to study planetesimal accretion and how magnetic fields influence the spreading of material across the atmosphere. To better study these processes however, more targets are needed. Here we report our findings from an observing campaign on the magnetic white dwarf LHS 2534. Time-resolved spectra at the MMT show clear variability in the strength of the \ion{Na}{1} and \ion{Cr}{1} features. We find that the strengths of these features vary over a period of 19.46 and 19.36 hours respectively, closely matching the 20.22 hour period determined from K2 photometry. Our findings demonstrate the ability to detect and study these variable objects with spectroscopy, and time-series observations of similar targets in the future will provide more insight into how these atmospheres form.

\end{abstract}


\section{Introduction}
The majority of white dwarfs are known to display stable, homogeneous atmospheres. However, some white dwarfs exhibit atmospheric inhomogeneities that can produce unusual spectral features. Metal-enriched polar caps, or ``metal scars'', have been reported in three magnetic, metal-polluted white dwarfs \citep{Bagnulo24.1,Bagnulo24.2,Bagnulo26}. As the white dwarf rotates, these surface inhomogeneities can produce periodic variations in spectral line strengths. The formation of these scars is likely due to the accretion of ionized gas, which is then guided along the field lines toward the poles. The magnetic field then likely prevents the spreading of this material across the surface \citep{Cunningham21,Dang26,Shiftan26}. These discoveries raise questions about the prevalence of metal scarring and spectral variability among DZH-class white dwarfs, as well as possible trends with rotation periods, ages, and field strengths. Here we report our findings from an observing campaign on the magnetic, metal-polluted white dwarf LHS 2534. This study presents the fourth known case of a magnetic, metal-polluted white dwarf exhibiting spectral variability. Because so few examples are known, identifying and characterizing additional objects is essential for determining the occurrence rate of this phenomenon and understanding the physical mechanisms responsible.

\section{Observations}

The majority of our data was taken at the 6.5m MMT with the Blue Channel Spectrograph. We used the 1.25" slit with the 500 l mm$^{-1}$ grating which yields a resolution of 4.5 {\AA}. We obtained three exposures per night over five nights for a total of 15 exposures.

We also obtained two exposures separated by two hours at the APO 3.5m telescope equipped with the KOSMOS spectrograph. We used the 2" slit in the Center position with the Blue disperser and 2x2 binning with a resolution of 4.7 {\AA}.

Finally, we obtained five back-to-back exposures and two exposures separated by 2.5 hours using the 8m Gemini North telescope with the Gemini Multi-Object Spectrograph as part of the program GS-2024B-Q-230. We used the B480 grating with the 1" slit and 4x4 binning, providing a resolution of 6.3 {\AA}. All of our spectra were reduced using standard IRAF procedures.

\section{Results {\&} Conclusion}
\begin{figure*}[ht!]
    \centering
    
    \subfloat[\label{fig:Comp}]{\includegraphics[width=0.48\textwidth]{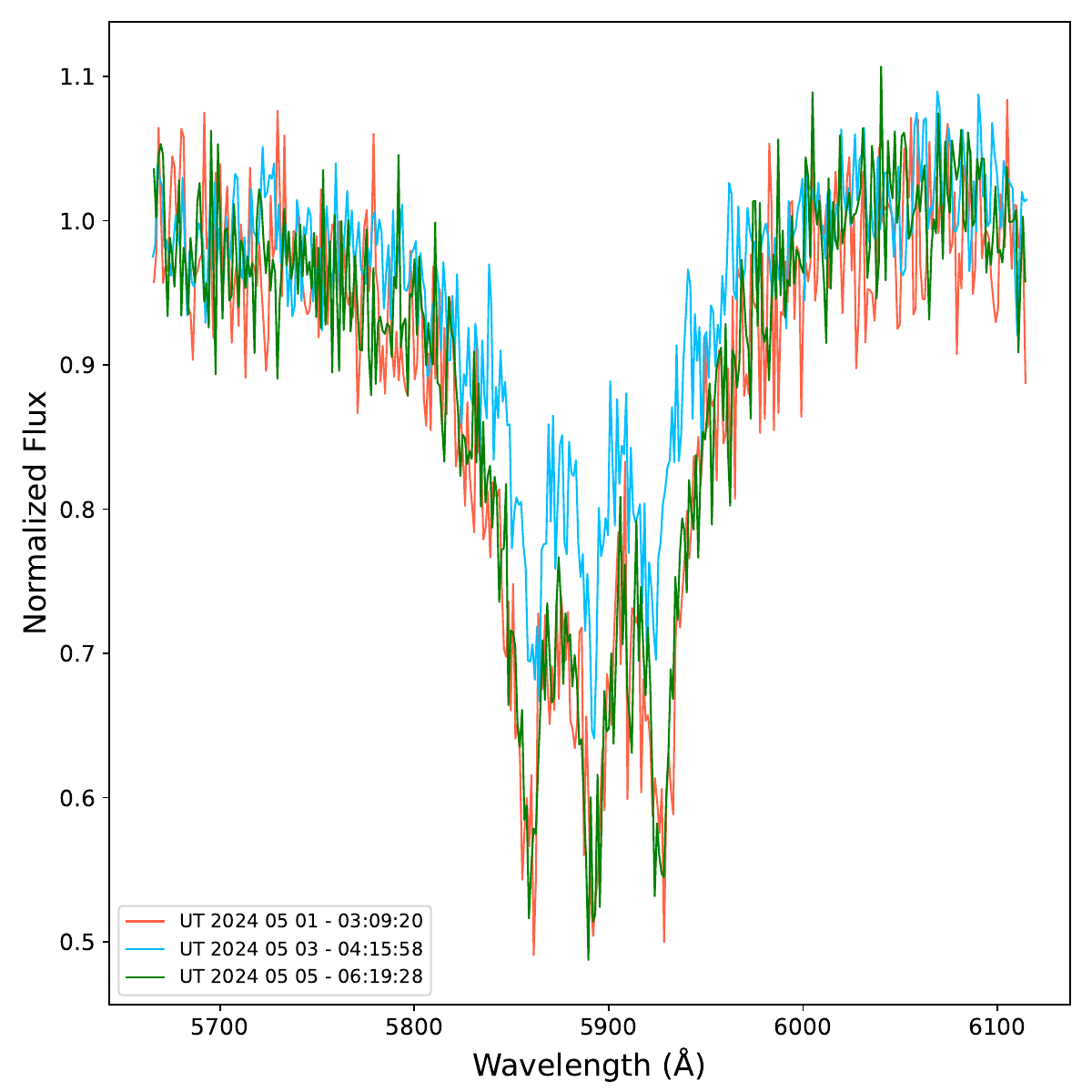}}
    \hspace{0.30cm}
    \subfloat[\label{fig:MMTN}]{\includegraphics[width=0.48\textwidth]{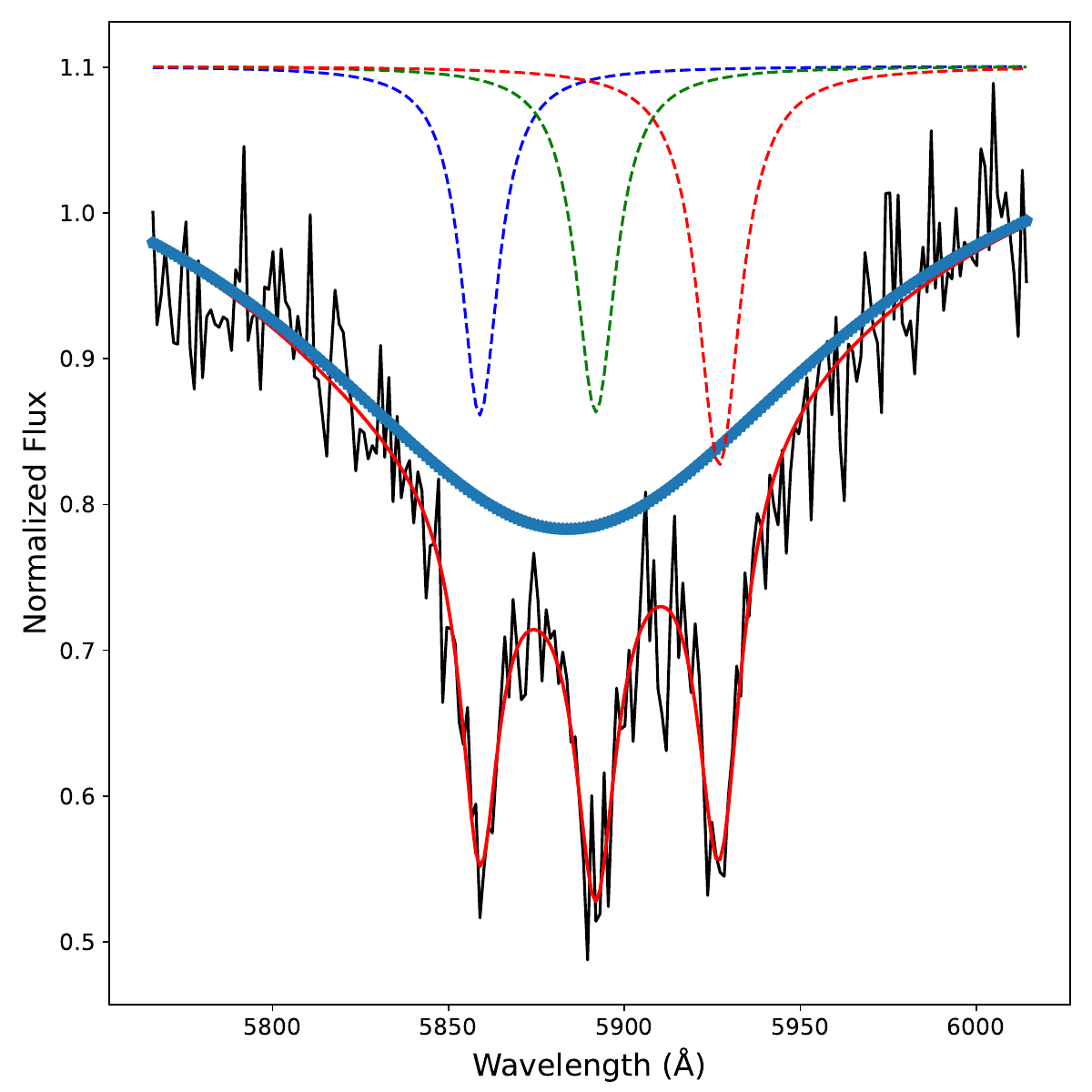}}
    
    \par\vspace{-0.30cm}
 
    \subfloat[\label{fig:MMTAD}]{\includegraphics[width=0.48\textwidth]{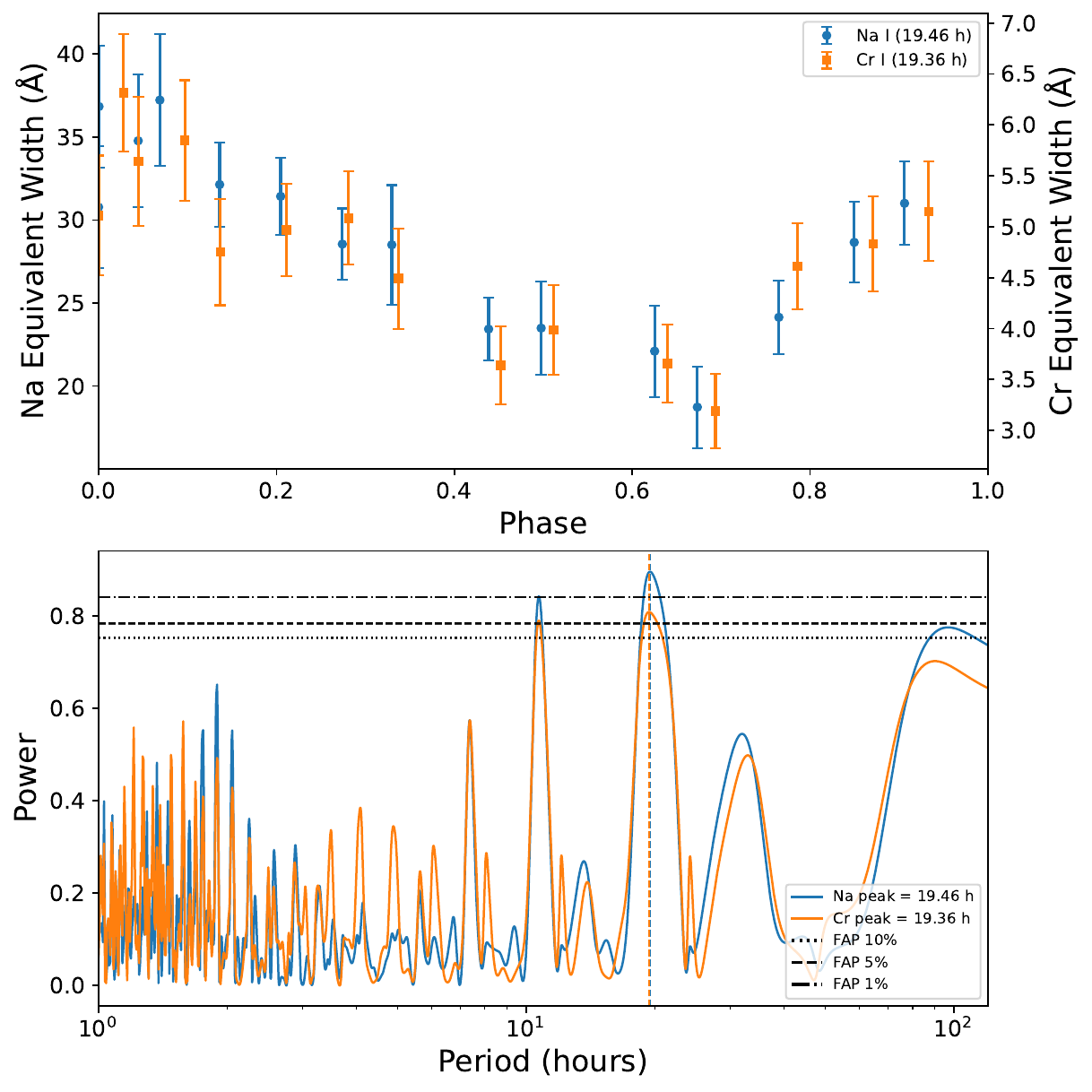}}
    \hspace{0.30cm}
    \subfloat[\label{fig:K2D}]{\includegraphics[width=0.48\textwidth]{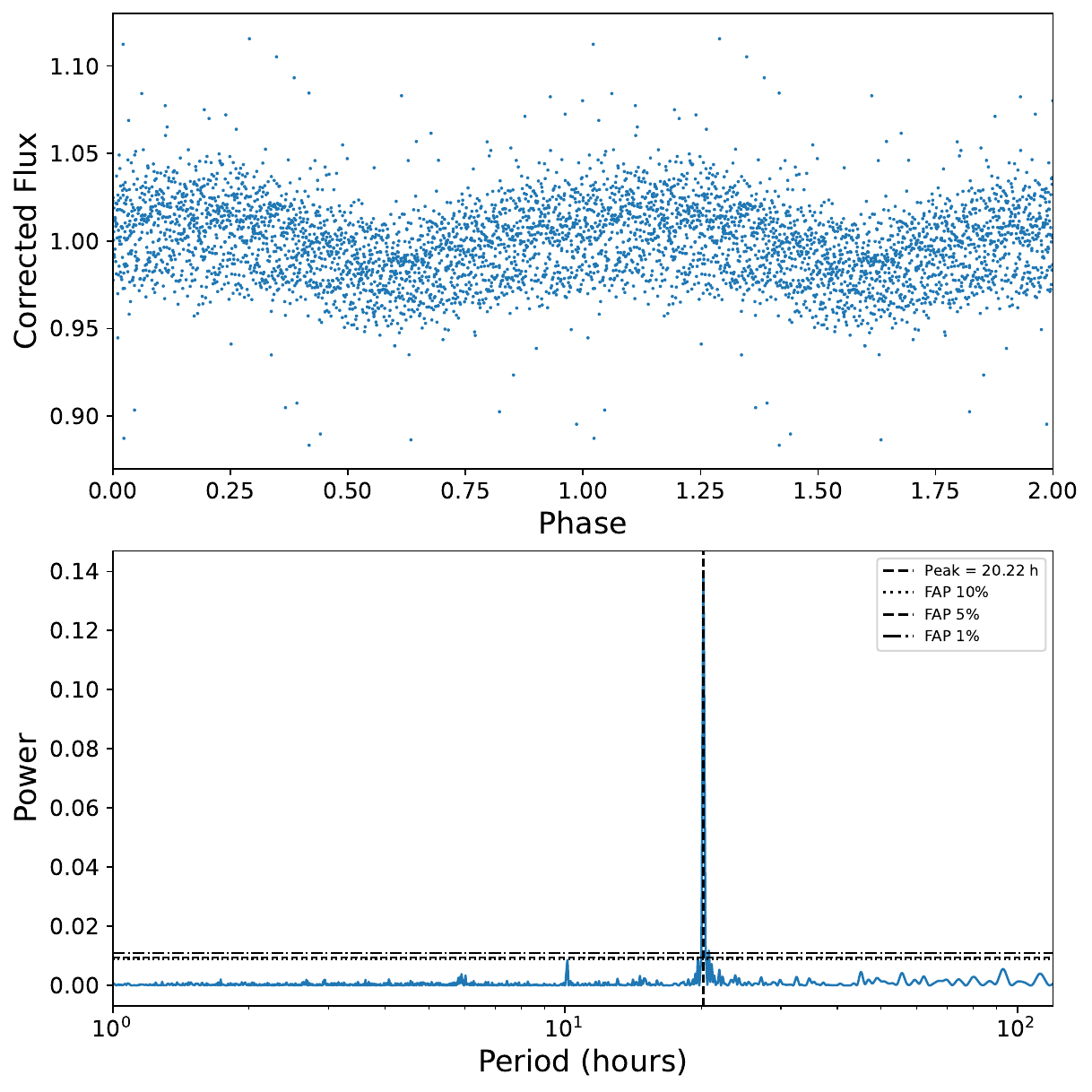}}

    \caption{(a) Comparison of normalized MMT spectra of LHS 2534, showing clear variability in the \ion{Na}{1} absorption profile. (b) Lorentzian fit to the \ion{Na}{1} absorption triplet, including the broad component used to account for its extended wings. (c) Phase-folded \ion{Na}{1} and \ion{Cr}{1} EWs from the MMT observations (top) and Lomb–Scargle periodograms of the MMT \ion{Na}{1} and \ion{Cr}{1} EW measurements (bottom). Horizontal lines indicate the 10\%, 5\%, and 1\% false-alarm probability thresholds. The primary peaks occur at 19.46 hours for \ion{Na}{1} and 19.36 hours for \ion{Cr}{1}. Secondary spectroscopic peaks occur near 10.69 hours. (d) The K2 light curve (top) phase-folded at the photometric period of 20.22 hours, and Lomb–Scargle periodogram of the K2 photometry (bottom).}

    \label{fig:combined}
    
\end{figure*}

The MMT spectra show clear changes in the \ion{Na}{1} absorption profile  (Figure~\ref{fig:combined}\subref*{fig:Comp}), demonstrating spectral variability in LHS 2534. To quantify this variability, we measured the equivalent widths (EWs) of the \ion{Na}{1} feature, which appears as a Zeeman-split triplet, and the \ion{Cr}{1} feature, which is likely blended with \ion{Mg}{1} \citep{Hollands21}. Following the spectral-fitting procedures described by \citet{Moss24,Moss25a}, we fit Lorentzian profiles to both features. To account for both the broad wings and narrow line cores, we included an additional broad component spanning each feature, which substantially improved the quality of the fits. An example of the \ion{Na}{1} fit is shown in Figure~\ref{fig:combined}\subref*{fig:MMTN}. We estimated the EW uncertainties by using bootstrapping to repeat the fitting procedure 10,000 times for each spectrum. The 50, 15.8, and 84.2 percentiles of each distribution provide the median EWs and the $\pm1\sigma$ lower and upper limits respectively. We also measured the EWs in the Gemini and APO spectra to search for variability. However, their limited phase coverage does not establish an independent periodic pattern, though one Gemini spectrum differs noticeably from the others.

We converted the observation times to Barycentric Julian Date (BJD) and calculated Lomb–Scargle periodograms using the MMT EW measurements. The top panel of Figure~\ref{fig:combined}\subref*{fig:MMTAD} shows the MMT EW measurements folded at the highest peaks in the Lomb-Scargle periodograms, which are shown in the bottom panel. The \ion{Na}{1} periodogram shows a primary peak at 19.46 hours and a secondary peak at approximately 10.70 hours. The \ion{Cr}{1} periodogram shows primary and secondary peaks at 19.36 and 10.69 hours, respectively. The agreement between the \ion{Na}{1} and \ion{Cr}{1} periods indicate rotation of an inhomogeneous atmosphere.

To determine if the rotation period from our observations is the true period, we analyzed the light curve of LHS 2534 from K2 photometry \citep{Vanderburg14}. We excluded 68 out of 2508 measurements that deviated by more than \(5\sigma\) from the median flux. The top panel of Figure~\ref{fig:combined}\subref*{fig:K2D} shows the K2 light curve folded at the strongest peak in the corresponding periodogram which is shown in the bottom panel. The K2 data shows clear photometric variations over a period of 20.22 hours. Both periods derived from the \ion{Na}{1} and \ion{Cr}{1} data closely match the true period from the K2 photometry given the phase coverage of our spectroscopy. Additional spectra over multiple nights with longer sequences would likely return the 20.22 hour period. 

Together, the variations in the \ion{Na}{1} and \ion{Cr}{1} features and the agreement between the K2 and MMT periods strongly suggest that LHS 2534 has a chemically inhomogeneous surface with a metal-enriched polar cap. Further observations of the DZH white dwarfs in the SDSS 100 pc sample \citep{Moss25b} could determine the prevalence of this behavior and investigate possible relationships with rotation period, magnetic field strength, and other stellar properties.
\\

We acknowledge support from the National Science Foundation under grant number 2348547 through the University of Florida 2026 REU program. 

Observations reported here were obtained at the MMT Observatory, a joint facility of the Smithsonian Institution and the University of Arizona. 

Based on observations obtained with the Apache Point Observatory 3.5-meter telescope, which is owned and operated by the Astrophysical Research Consortium. 

Based on observations obtained at the international Gemini Observatory, a program of NSF’s NOIRLab, which is managed by the Association of Universities for Research in Astronomy (AURA) under a cooperative agreement with the National Science Foundation on behalf of the Gemini Observatory partnership: the National Science Foundation (United States), National Research Council (Canada), Agencia Nacional de Investigación y Desarrollo (Chile), Ministerio de Ciencia, Tecnología e Innovación (Argentina), Ministério da Ciência, Tecnologia, Inovações e Comunicações (Brazil), and Korea Astronomy and Space Science Institute (Republic of Korea).

\bibliography{Main}{}

@ARTICLE{Moss25a,
       author = {{Moss}, Adam and {Kilic}, Mukremin and {Bergeron}, Pierre and {Jewett}, Gracyn and {Brown}, Warren R.},
        title = "{The Emerging Class of Double-faced White Dwarfs}",
      journal = {\apj},
         year = 2025,
        month = apr,
       volume = {983},
       number = {1},
          eid = {14},
        pages = {14},
          doi = {10.3847/1538-4357/adbd3a},
archivePrefix = {arXiv},
       eprint = {2501.05649},
 primaryClass = {astro-ph.SR},
       adsurl = {https://ui.adsabs.harvard.edu/abs/2025ApJ...983...14M}
}

@ARTICLE{Moss25b,
       author = {{Moss}, Adam and {Kilic}, Mukremin and {Bergeron}, Pierre and {Brown}, Warren R. and {Jewett}, Gracyn and {Ag{\"u}eros}, Marcel A. and {Camisassa}, Maria and {Burrow}, Anthony},
        title = "{Magnetic White Dwarfs in the SDSS 100 pc Sample: Further Evidence of Two Formation Channels}",
      journal = {\apj},
         year = 2025,
        month = sep,
       volume = {990},
       number = {1},
          eid = {25},
        pages = {25},
          doi = {10.3847/1538-4357/aded8f},
archivePrefix = {arXiv},
       eprint = {2507.06102},
 primaryClass = {astro-ph.SR},
       adsurl = {https://ui.adsabs.harvard.edu/abs/2025ApJ...990...25M}
}

@ARTICLE{Bagnulo24.1,
       author = {{Bagnulo}, Stefano and {Farihi}, Jay and {Landstreet}, John D. and {Folsom}, Colin P.},
        title = "{Discovery of Magnetically Guided Metal Accretion onto a Polluted White Dwarf}",
      journal = {\apjl},
         year = 2024,
        month = mar,
       volume = {963},
       number = {1},
          eid = {L22},
        pages = {L22},
          doi = {10.3847/2041-8213/ad2619},
archivePrefix = {arXiv},
       eprint = {2402.16526},
 primaryClass = {astro-ph.SR},
       adsurl = {https://ui.adsabs.harvard.edu/abs/2024ApJ...963L..22B}
}

@ARTICLE{Moss24,
       author = {{Moss}, Adam and {Bergeron}, P. and {Kilic}, Mukremin and {Jewett}, Gracyn and {Brown}, Warren R. and {Kosakowski}, Alekzander and {Vincent}, Olivier},
        title = "{Discovery of a magnetic double-faced DBA white dwarf}",
      journal = {\mnras},
         year = 2024,
        month = feb,
       volume = {527},
       number = {4},
        pages = {10111-10122},
          doi = {10.1093/mnras/stad3825},
       adsurl = {https://ui.adsabs.harvard.edu/abs/2024MNRAS.52710111M}
}

@ARTICLE{Bagnulo24.2,
       author = {{Bagnulo}, S. and {Landstreet}, J.~D. and {Farihi}, J. and {Folsom}, C.~P. and {Hollands}, M.~A. and {Fossati}, L.},
        title = "{Metal accretion scars may be common on magnetic, polluted white dwarfs}",
      journal = {\aap},
         year = 2024,
        month = aug,
       volume = {688},
          eid = {L14},
        pages = {L14},
          doi = {10.1051/0004-6361/202451215},
archivePrefix = {arXiv},
       eprint = {2407.17196},
 primaryClass = {astro-ph.SR},
       adsurl = {https://ui.adsabs.harvard.edu/abs/2024A&A...688L..14B}
}

@ARTICLE{Hollands21,
       author = {{Hollands}, Mark A. and {Tremblay}, Pier-Emmanuel and {G{\"a}nsicke}, Boris T. and {Koester}, Detlev and {Gentile-Fusillo}, Nicola Pietro},
        title = "{Alkali metals in white dwarf atmospheres as tracers of ancient planetary crusts}",
      journal = {Nature Astronomy},
         year = 2021,
        month = may,
       volume = {5},
        pages = {451-459},
          doi = {10.1038/s41550-020-01296-7},
archivePrefix = {arXiv},
       eprint = {2101.01225},
 primaryClass = {astro-ph.EP},
       adsurl = {https://ui.adsabs.harvard.edu/abs/2021NatAs...5..451H}
}

@ARTICLE{Cunningham21,
       author = {{Cunningham}, Tim and {Tremblay}, Pier-Emmanuel and {Bauer}, Evan B. and {Toloza}, Odette and {Cukanovaite}, Elena and {Koester}, Detlev and {Farihi}, Jay and {Freytag}, Bernd and {G{\"a}nsicke}, Boris T. and {Ludwig}, Hans-G{\"u}nter and {Veras}, Dimitri},
        title = "{Horizontal spreading of planetary debris accreted by white dwarfs}",
      journal = {\mnras},
         year = 2021,
        month = may,
       volume = {503},
       number = {2},
        pages = {1646-1667},
          doi = {10.1093/mnras/stab553},
archivePrefix = {arXiv},
       eprint = {2102.09564},
 primaryClass = {astro-ph.SR},
       adsurl = {https://ui.adsabs.harvard.edu/abs/2021MNRAS.503.1646C}
}

@ARTICLE{Bagnulo26,
       author = {{Bagnulo}, S. and {Stroet}, M.~A. and {Folsom}, C.~P. and {Hollands}, M.~A. and {Landstreet}, J.~D. and {Ramsay}, G.},
        title = "{Magnetically guided accretion and extremely slow rotation in a metal-enriched white dwarf}",
      journal = {\aap},
         year = 2026,
        month = sep,
       volume = {713},
          eid = {A220},
        pages = {A220},
          doi = {10.1051/0004-6361/202660714},
archivePrefix = {arXiv},
       eprint = {2607.21846},
 primaryClass = {astro-ph.SR},
       adsurl = {https://ui.adsabs.harvard.edu/abs/2026A&A...713A.220B}
}

@ARTICLE{Dang26,
       author = {{Pham}, Dang and {Taylor}, Aster G. and {Cunningham}, Tim},
        title = "{The Effects of Magnetic Accretion on the Spatial Extent of White Dwarf Pollution}",
      journal = {\apj},
         year = 2026,
        month = sep,
       volume = {1008},
       number = {2},
          eid = {257},
        pages = {257},
          doi = {10.3847/1538-4357/ae8fd5},
archivePrefix = {arXiv},
       eprint = {2607.20747},
 primaryClass = {astro-ph.SR},
       adsurl = {https://ui.adsabs.harvard.edu/abs/2026ApJ..1008..257P}
}

@ARTICLE{Shiftan26,
       author = {{Shiftan}, Elad and {Ginzburg}, Sivan},
        title = "{Sinking and spreading of metal pollution in magnetic white dwarfs}",
      journal = {arXiv e-prints},
         year = 2026,
        month = aug,
          eid = {arXiv:2609.00117},
        pages = {arXiv:2609.00117},
          doi = {10.48550/arXiv.2609.00117},
archivePrefix = {arXiv},
       eprint = {2609.00117},
 primaryClass = {astro-ph.SR},
       adsurl = {https://ui.adsabs.harvard.edu/abs/2026arXiv260900117S}
}

@ARTICLE{Vanderburg14,
       author = {{Vanderburg}, Andrew and {Johnson}, John Asher},
        title = "{A Technique for Extracting Highly Precise Photometry for the Two-Wheeled Kepler Mission}",
      journal = {\pasp},
         year = 2014,
        month = oct,
       volume = {126},
       number = {944},
        pages = {948},
          doi = {10.1086/678764},
archivePrefix = {arXiv},
       eprint = {1408.3853},
 primaryClass = {astro-ph.IM},
       adsurl = {https://ui.adsabs.harvard.edu/abs/2014PASP..126..948V}
}
\bibliographystyle{aasjournal7.1}

\end{document}